\documentclass[runningheads,a4paper]{llncs}

\usepackage{amssymb}
\usepackage{graphicx}
  \usepackage{color}
  
    \usepackage{float}
    \usepackage{wrapfig}
\usepackage{subfigure}

\usepackage{url}
\urldef{\mailsa}\path|{alfred.hofmann, ursula.barth, ingrid.haas, frank.holzwarth,|
\urldef{\mailsb}\path|anna.kramer, leonie.kunz, christine.reiss, nicole.sator,|
\urldef{\mailsc}\path|erika.siebert-cole, peter.strasser, lncs}@springer.com|

\begin{document}

\mainmatter  

%
%
%
%
%

\title{Context Aware Sensor Configuration Model for Internet of Things}


\titlerunning{Semantic-driven Configuration in IoT Middleware}

%
%
\author{Charith Perera\inst{1,2} \and 
Arkady Zaslavsky\inst{2} \and 
Michael Compton\inst{2} \and Peter Christen\inst{1} \and Dimitrios Georgakopoulos\inst{2}}
\authorrunning{Perera et al.}   
%
\tocauthor{Charith Perera, Arkady Zaslavsky, Michael Compton, Peter Christen, Dimitrios Georgakopoulos}
\institute{Research School of Computer Science, The Australian National University, Canberra, ACT 0200, Australia\\
\email{\{charith.perera, peter.christen\}@anu.edu.au},\\ 
\and
CSIRO ICT Center, Canberra, ACT 2601, Australia\\
\email{\{charith.perera, arkady.zaslavsky, michael.compton dimitrios.georgakopoulos\}@csiro.au}
}

%
%

\toctitle{Lecture Notes in Computer Science}
\tocauthor{Authors' Instructions}
\maketitle

\begin{abstract}

We propose a \textit{Context Aware Sensor Configuration Model} (CASCoM) to address the challenge of automated context-aware configuration of filtering, fusion, and reasoning mechanisms in IoT middleware according to the problems at hand. We incorporate semantic technologies in solving the above challenges.



\end{abstract}

\vspace{-25pt}

\section{Introduction}
\label{sec:Introduction}

 Broadly, configuration in IoT paradigm can be categorized into two: \textit{sensor-level} configuration and \textit{system-level} configuration. Sensor-level configuration focuses on changing a sensor's behaviour by configuring  its embedded software parameters such as sensing schedule, sampling rate, data communication frequency, communication patterns and protocols. In this paper, we are focused on developing a system-level configuration model for IoT midddleware platforms. Specifically, our proposed model identifies, composes, and configures both sensors and data processing components according to the user requirements. In existing IoT middleware (e.g. GSN), many configuration files and programming codes need to be manually defined by the users (without any help from GSN). An ideal IoT middleware configuration model should address all the above mentioned challenges. The configuration model we propose in this paper is applicable towards several other emerging paradigms, such as sensing as a service \cite{ZMP003}.

\vspace{-8pt}

\section{Problem Analysis}
\label{sec:Problem_Definition}

\vspace{-8pt}



\begin{figure}[h]
 \centering
 \includegraphics[scale=.75]{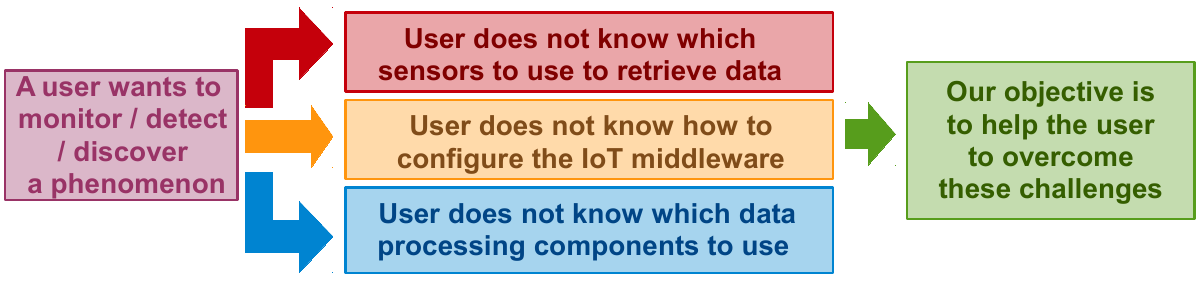}
\vspace{-0.35cm}	
 \caption{Research Challenges, User Requirements, and Our Objective}
 \label{Figure:Problem_Analysis_General}	
\vspace{-0.33cm}	
\end{figure}

\begin{figure}[h]
 \centering
 \includegraphics[scale=.70]{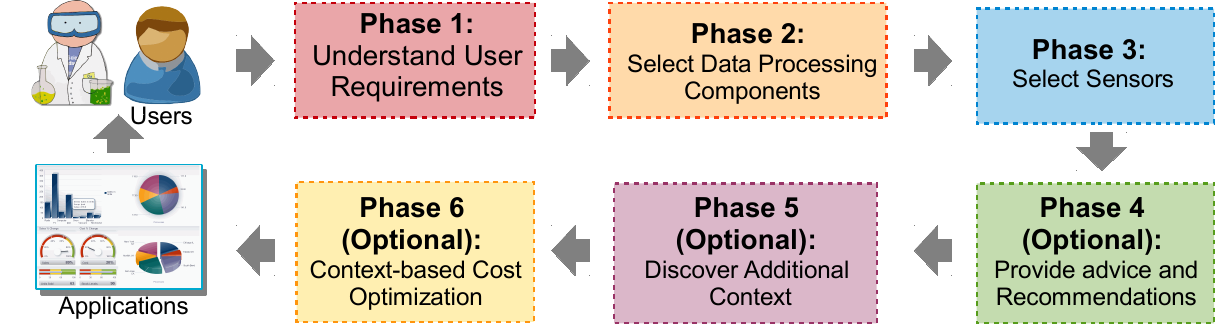}
\vspace{-0.43cm}	
 \caption{The Context-Aware Sensor Configuration Model (CASCoM)}
 \label{Figure:The_Model}	
\end{figure}

Our research question is \textit{`How to develop a model that allows non-IT experts to configure sensors and data processing mechanisms in an IoT middleware according to the user requirements?'}. Extended explanations are provided in \cite{ZMP004}. Research challenges are highlighted in Figure \ref{Figure:Problem_Analysis_General}. Context-Aware Sensor Configuration Model (CASCoM) simplifies the IoT middleware configuration process significantly. Figure \ref{Figure:configuration_Workflow_Comparison} compares the execution-flow of sensor configuration in the current GSN approach and the CASCoM approach. The proposed solution CASCoM is illustrated in Figure \ref{Figure:The_Model}.

\section{The CASCoM Architecture}
\label{sec:Proposed_Solution}

\vspace{-8pt}


\begin{figure}[b]
 \centering
\vspace{-0.63cm}
 \includegraphics[scale=.78]{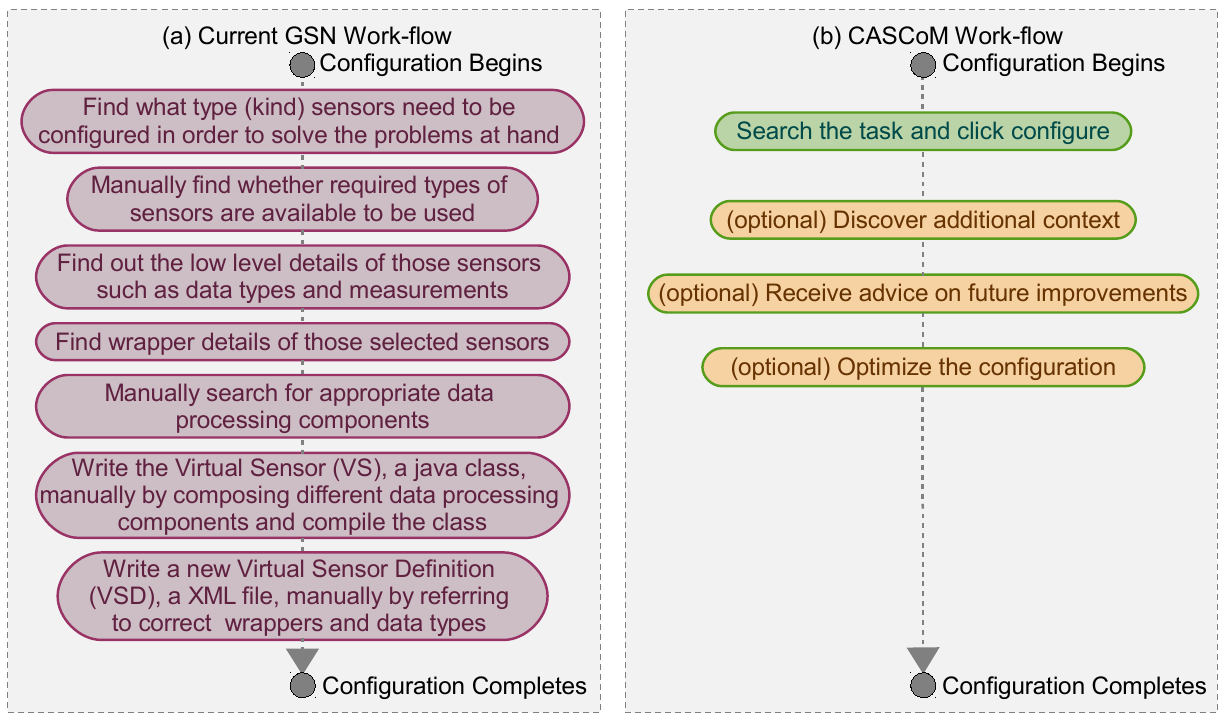}
\vspace{-0.43cm}	
 \caption{Configuration Execution-flow Comparison: (a) Current GSN  (b) CASCoM}
 \label{Figure:configuration_Workflow_Comparison}	
\end{figure}


In phase 1, users are facilitated with a graphical user interface, which is based on a \textit{question-answer (QA)} approach, that allows to express the user requirements. Users can answer as many question as possible. CASCoM searches and filters the tasks that the user may want to perform. From the filtered list, users can select the desired task. The details of the QA approach are presented later in this section. In phase 2, CASCoM searches for different programming components that allow to generate the data stream required. In phase 3, CASCoM tries to find the sensors that can be used to produce the inputs required by the selected data processing components. If CASCoM fails to produce the data streams required by the users due to insufficient resources (i.e. unavailability of the sensors), it will provide advice and recommendations on future sensor deployments in phase 4. Phase 5 allows the users to capture additional context information. The additional context information that can be derived using available resources and knowledge are listed to be selected.  In phase 6, users are provided with one or more solutions\footnote{Solution is a combination of sensors and data processing components that can be composed together in order to satisfy the user requirements.}. CASCoM calculates the costs for each solution in using technique disucced in \cite{ZMP006}. By default, CASCoM will select the solution with lowest cost. However, users can select the cost models as they required. Finally, CASCoM generates all the configuration files and program codes which we listed in Figure \ref{Figure:configuration_Workflow_Comparison}(a). Data starts streaming soon after.

\begin{figure}[h]
 \centering
 \includegraphics[scale=.56]{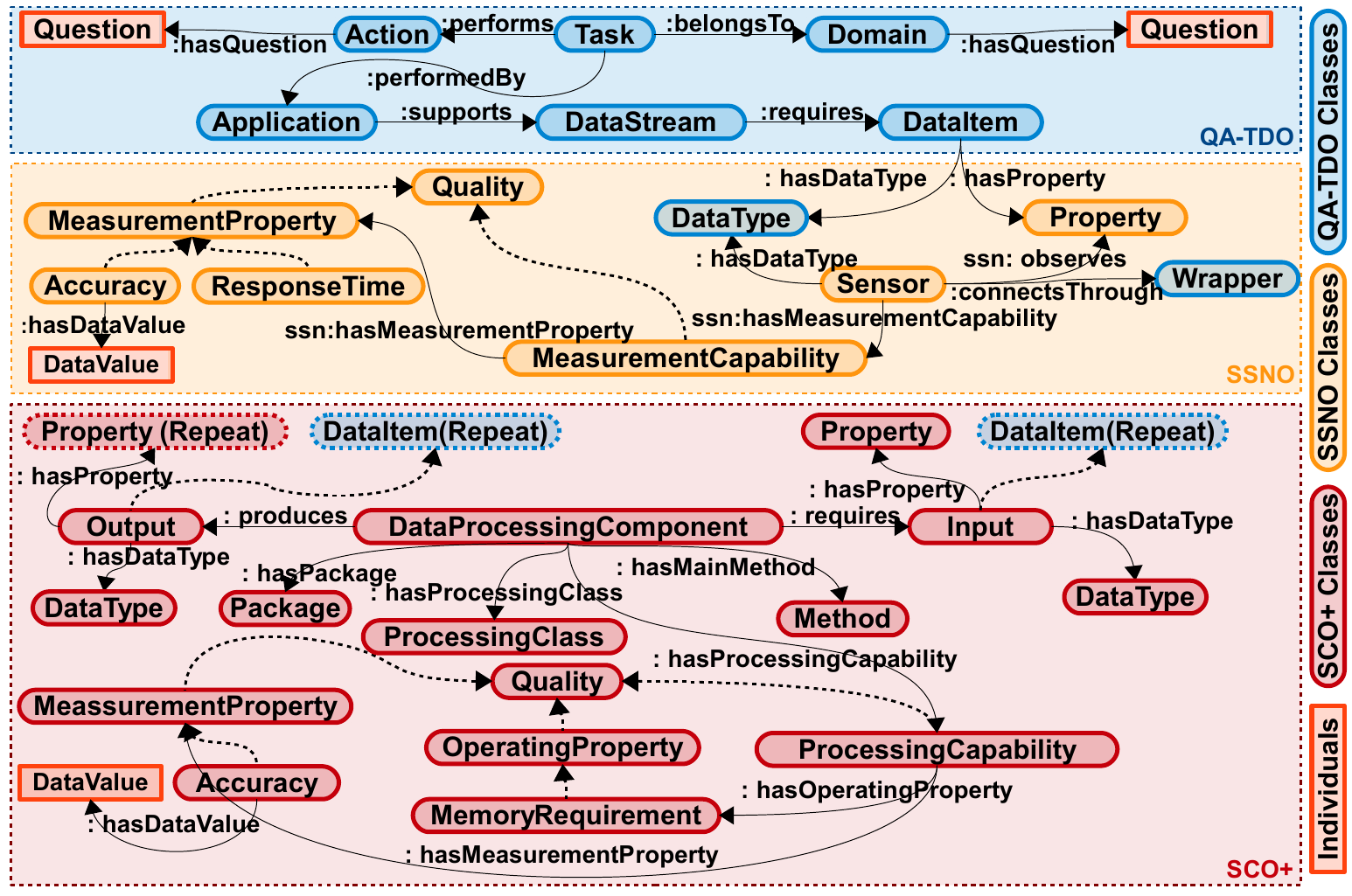}
\vspace{-0.43cm}	
 \caption{Extracts of different ontological data models used in CASCoM: QA-TDO, SCO \cite{P612}, and SSN ontology (w3.org/2005/Incubator/ssn/wiki/SSN). These model are used to store sensor descriptions, software component description, and domain knowledge. 
}
 \label{Figure:Ontology_Model}	
\vspace{-0.53cm}	
\end{figure}

\vspace{-10pt}

\vspace{-8pt}

\section{Evaluation, Discussion and Lessons Learned}
\label{sec:Discussion}

\begin{figure}[h]
\centering
\mbox{\subfigure{ \includegraphics[scale=0.36]{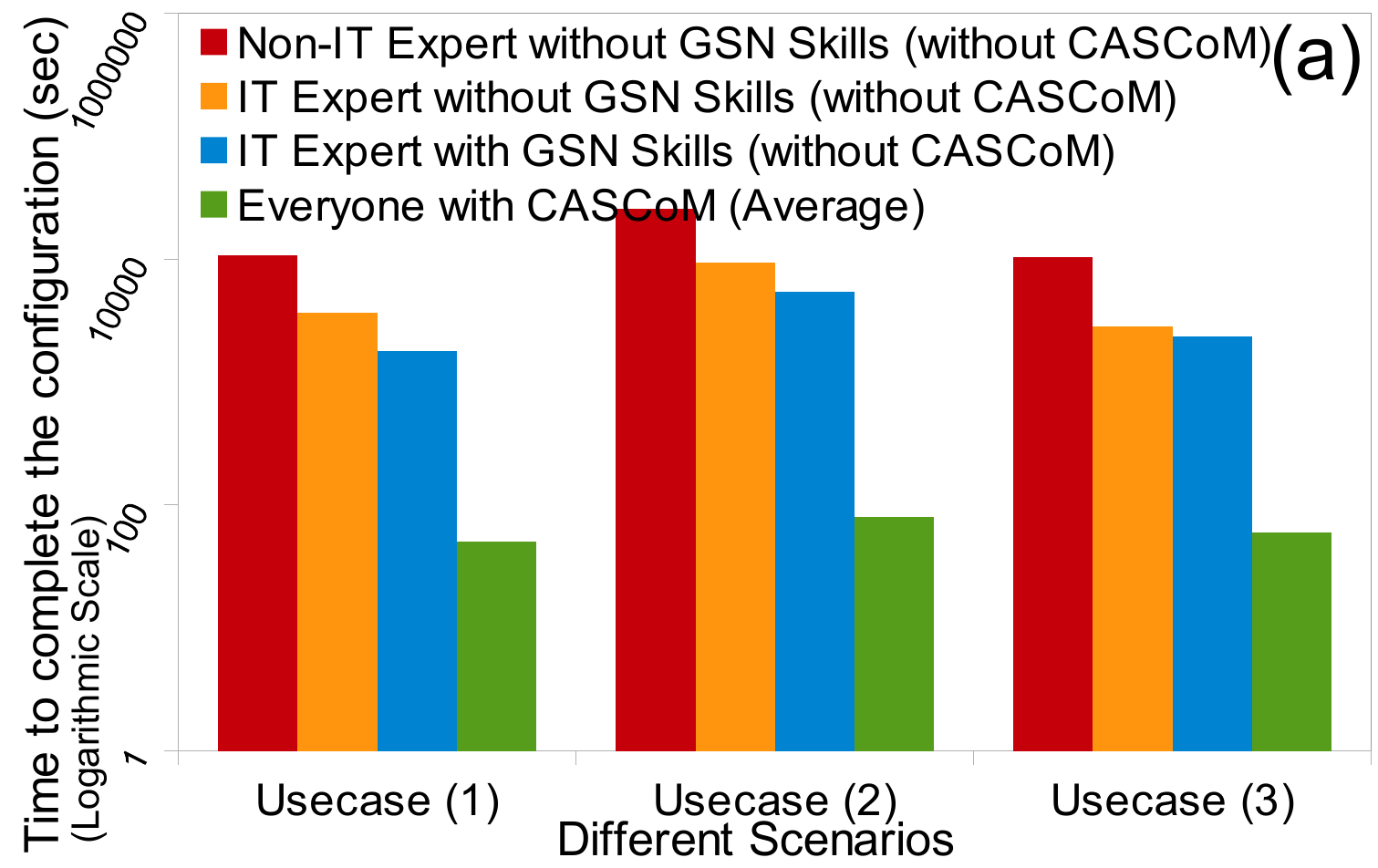}

\quad
\subfigure{ \includegraphics[scale=0.36]{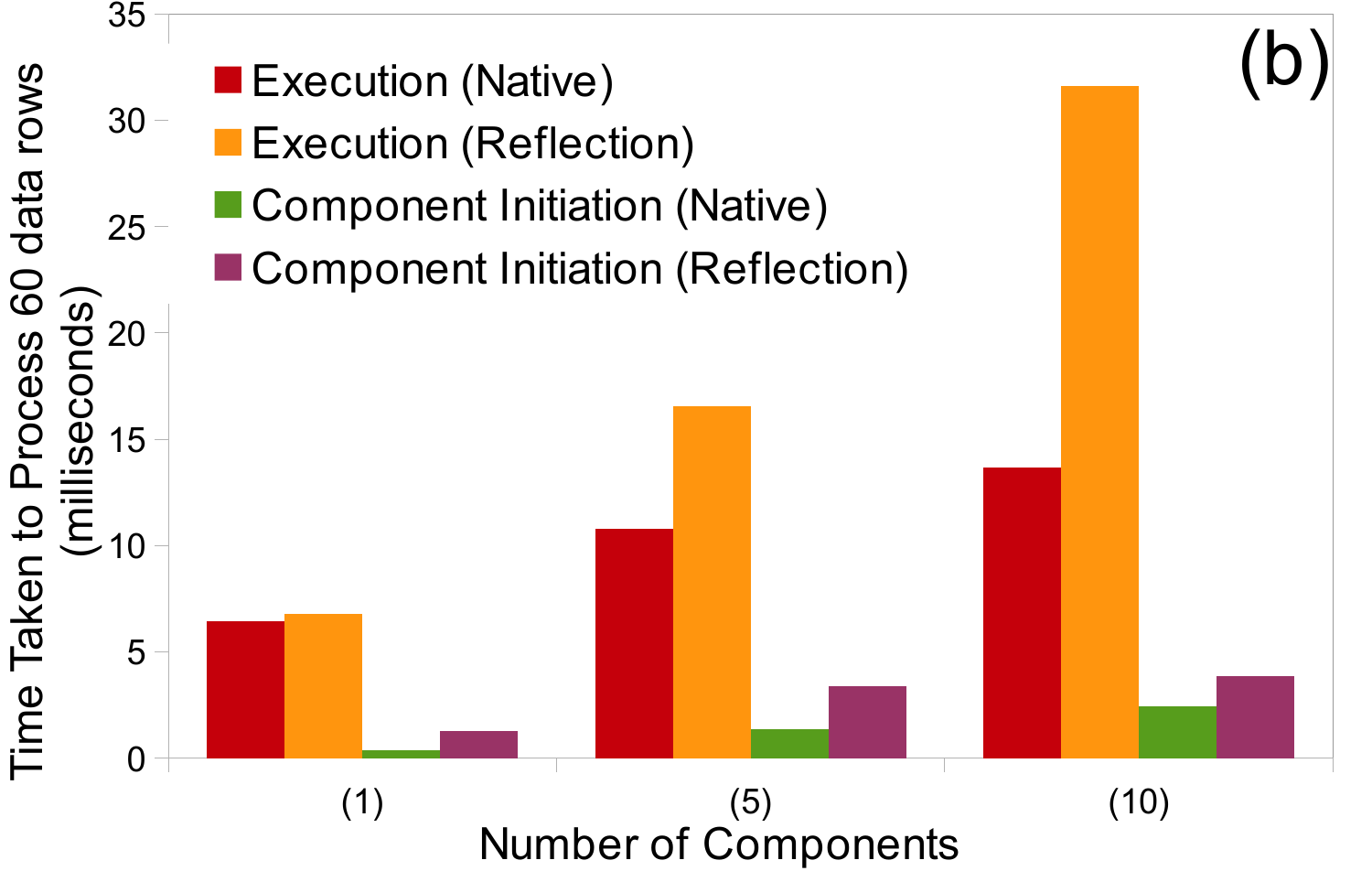} }
} }

\mbox{ \subfigure{ \includegraphics[scale=0.36]{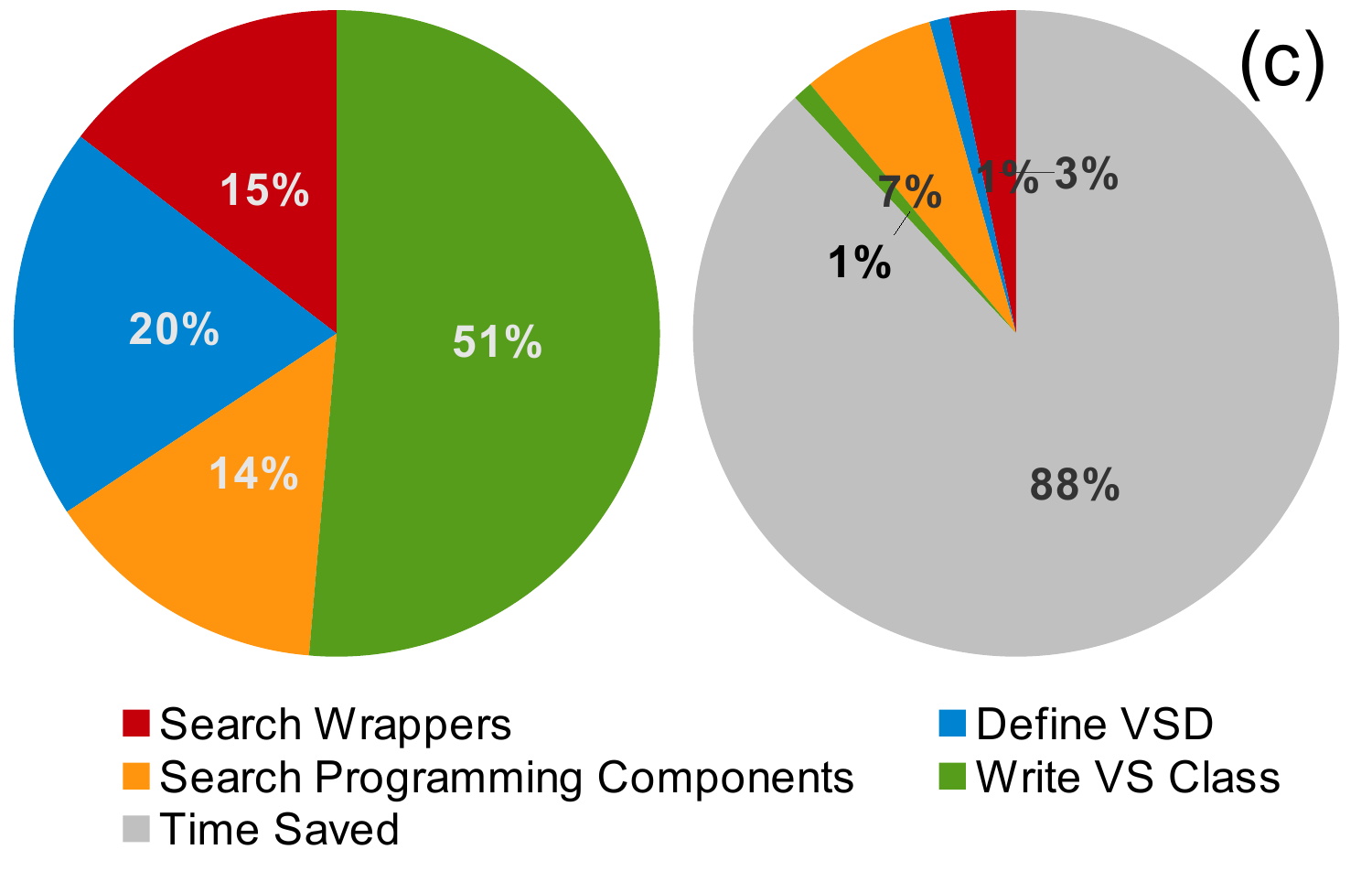}

\quad
\subfigure{ \includegraphics[scale=0.36]{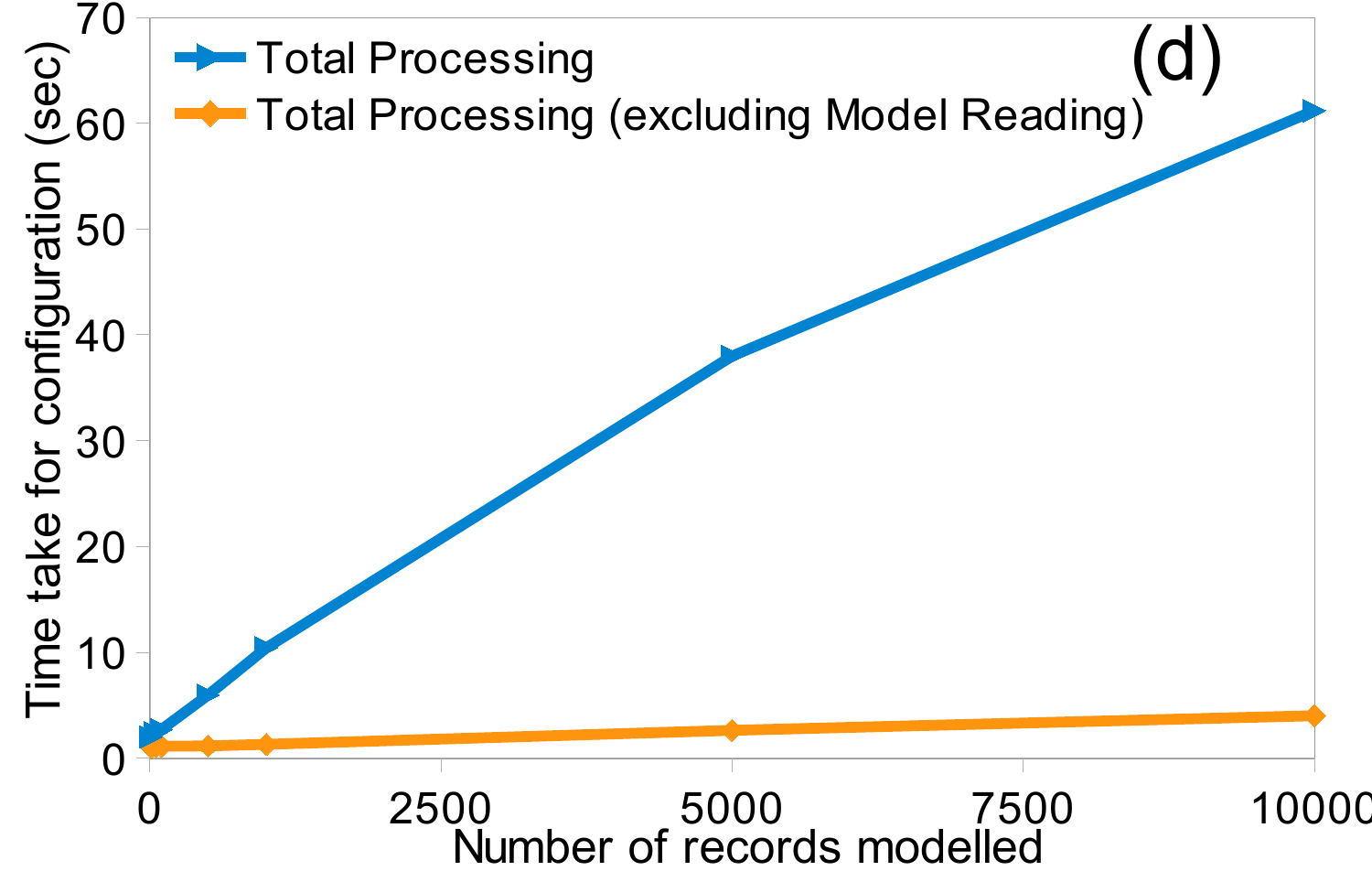} }
}}

\vspace{-20pt}
\caption{Evaluation of CASCoM}
\vspace{-15pt}
\label{Figure:Results}
\end{figure}

\textbf{Results:} Figure \ref{Figure:Results}(a) shows that CASCoM allows to considerably reduce the time required for configuration of data processing mechanism in IoT middleware. Specifically, CASCoM allowed the three types of users to complete the given task 50, 80 and 250 times faster (respectively) in comparison to the existing approach. \textcolor{black}{According to Figure \ref{Figure:Results}(b), the Java reflection approach takes slightly more time to specially when initializing. Though the Java reflection approach can add more flexibility to our model, the additional overhead increases when the number of components and operation involved gets increased. The overheads can grow up to unacceptable level very quickly when GSN scales up (e.g. more user requests).}

According to Figure \ref{Figure:Results}(c), even  IT experts who know GSN can save time by using CASCoM up to 88\%. Specially, time taken for defining the VSD and VS class have been significantly reduced. Both files can be generated by CASCoM autonomously within a second even for complex scenarios. However, the time taken to find data processing components and sensors (and wrappers) depends on the size of the semantic data model. Figure \ref{Figure:Results}(d) shows how total processing time would vary depending on the size of the semantic data model. Approximately, a semantic model with 10,000 sensor descriptions and 10,000 data processing components can be processed in order to find solutions for a given user request in less than a minute. However, most of the time is taken to read the data model. The actual configuration process other than reading the data model takes only 4 seconds and it slightly increases when the model size increases.

\vspace{-6pt}

\section{Conclusion}
\label{sec:Conclusion}

\vspace{-6pt}
%

We have shown that it is possible to offer a sophisticated configuration model to support non-IT experts. Semantic technologies are used extensively to support this model.  Using our proof of concept implementation, both IT and non-IT experts were able to configure the GSN in significantly less time. In future, we plan to extend our configuration model into sensor-level. To achieve this, we will develop a model that can be used to configure sensors autonomously without human intervention in highly dynamic smart environments in the IoT paradigm.



\vspace{-11pt}







  \bibliography{Bibliography}
  \bibliographystyle{abbrv}

\end{document}